# Deep Neural Networks Carve the Brain at its Joints

Maxwell A. Bertolero[1], Dustin Moraczewski[2], Adam Thomas[2], & Danielle S. Bassett[3-8]

[1]Department of Bioengineering, School of Engineering & Applied Science, University of Pennsylvania, Philadelphia, PA 19104 USA
[2]Data Sharing and Science Team, National Institute of Mental Health, Bethesda, MD, United States
[3]Department of Psychiatry, Perelman School of Medicine, University of Pennsylvania, Philadelphia, PA 19104 USA
[4]Department of Electrical & Systems Engineering, School of Engineering & Applied Science, University of Pennsylvania, Philadelphia, PA, 19104 USA
[5]Department of Neurology, Perelman School of Medicine, University of Pennsylvania, Philadelphia, PA, 19104 USA
[6]Department of Physics & Astronomy, College of Arts & Sciences, University of Pennsylvania, Philadelphia, PA, 19104 USA
[7]Santa Fe Institute, Santa Fe, NM, 87501 USA
[8]To whom correspondence should be address: dsb@seas.upenn.edu

**Abstract**
How an individual's unique brain connectivity determines that individual's cognition, behavior, and risk for pathology is a fundamental question in basic and clinical neuroscience. In seeking answers, many have turned to machine learning, with some noting the particular promise of deep neural networks in modelling complex non-linear functions. However, it is not clear that complex functions actually exist between brain connectivity and behavior, and thus if deep neural networks necessarily outperform simpler linear models, or if their results would be interpretable. Here we show that, across 52 subject measures of cognition and behavior, deep neural networks fit to each brain region's connectivity outperform linear regression, particularly for the brain's connector hubs—regions with diverse brain connectivity—whereas the two approaches perform similarly when fit to brain systems. Critically, averaging deep neural network predictions across brain regions results in the most accurate predictions, demonstrating the ability of deep neural networks to easily model the various functions that exists between regional brain connectivity and behavior, carving the brain at its joints. Finally, we shine light into the black box of deep neural networks using multislice network models. We determined that the relationship between connector hubs and behavior is best captured by modular deep neural networks. Our results demonstrate that both simple and complex relationships exist between brain connectivity and behavior, and that deep neural networks can fit both. Moreover, deep neural networks are particularly powerful when they are first fit to the various functions of a system independently and then combined. Finally, deep neural networks are interpretable when their architectures are structurally characterized using multislice network models.

**Main**
Human cognitive neuroscience seeks to explain how the function of an individual's brain determines their behavior. The human brain's functional connectivity, as commonly measured by the pairwise Pearson's correlation coefficient between regional time series, has generated tremendous insight into the brain's network function[1–19]. A desire to leverage individual variability in these connections to predict cognition, behavior, and symptoms of mental illness has united neuroscientists, clinicians, and machine learning experts[20–32]. Yet, precisely which prediction algorithm will prove most efficacious is unknown. While deep neural networks display notable predictive power in other domains[33], it is not clear whether their application to human brain connectivity is necessary or insightful[34]. Further, it is not known whether the mathematical functions that define the relationship between brain connectivity and a given outcome are sufficiently complex so as to require deep neural networks instead of simpler linear models[34,35]. Moreover, even should deep neural networks offer accurate predictions, their interpretability has been rightfully questioned[20,36–38].

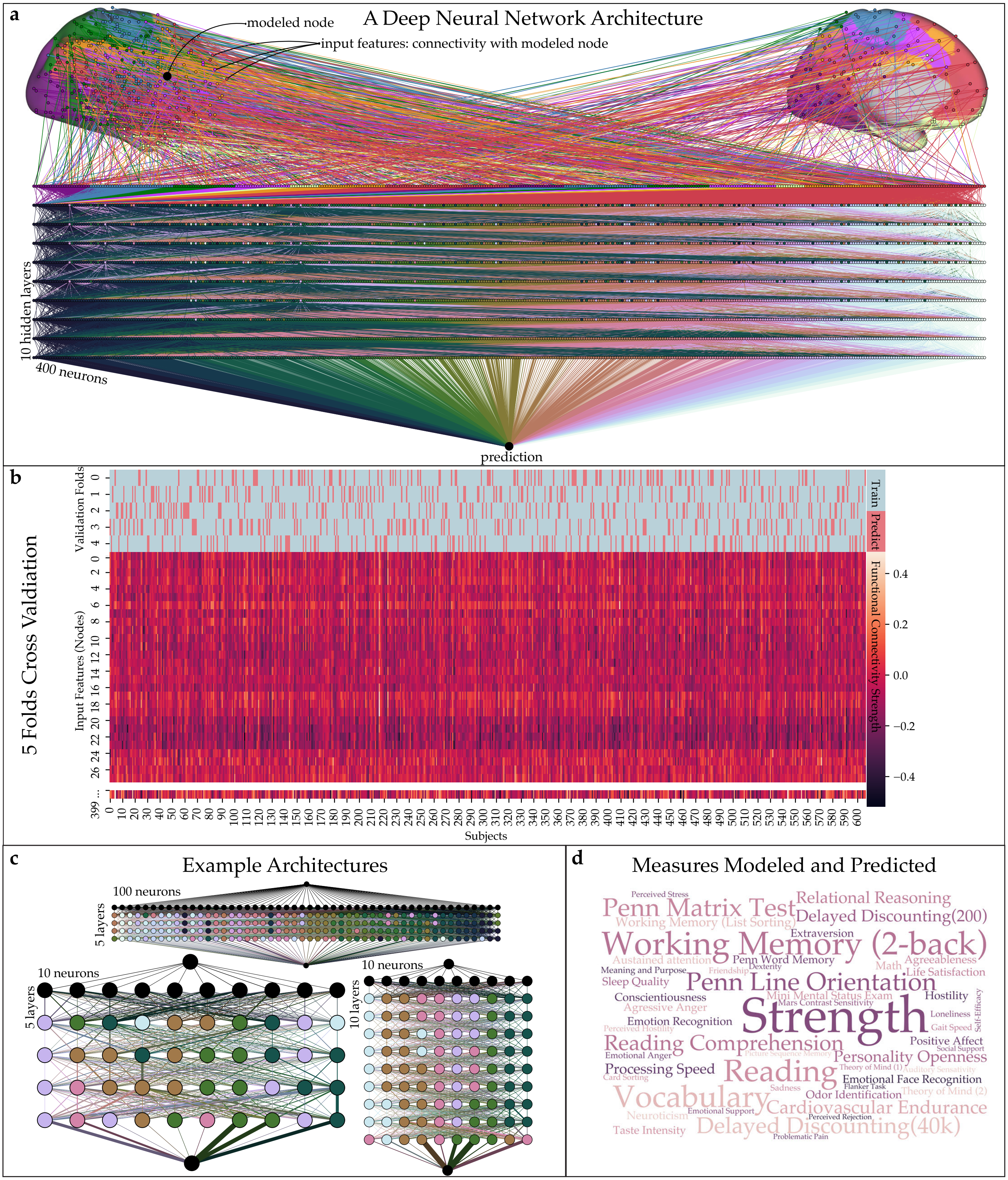

Figure 1 | **Using deep neural networks to predict behavior**. **a**, For each region in the brain, we used that region's functional connectivity (to *n*=399 other regions) to predict a given subject measure (e.g., working memory performance). This image was generated with Gephi[39]. The 399 connectivity weights are imputed into a deep neural network, with, for example, 10 hidden layers and 400 neurons. The weights between neurons in adjacent layers are trained to accurately relate variance across subjects in the modeled region's functional connectivity weights to variance in the subject measure across subjects. Regions are colored according to their community affiliation (Methods). **b**, We used 5-fold cross validation, where the subjects are split into five groups, the model is trained on 80 percent of the subjects (i.e., four groups), and then the model is used to predict the subject measure on the 20 percent of subjects that the model has not

seen. **c**, We tested multiple deep neural network architectures, composed of 1 to 10 hidden layers and 10 to 400 neurons. Nodes in the deep neural networks are colored according to their community affiliation (Methods). **d**, We modeled and predicted 52 subject measures, from working memory to social cognition to odor identification. Here, darker and larger text represents a higher prediction accuracy in deep neural network models.

Here, we use linear regression, support vector machines, and deep neural networks to iteratively predict 52 subject measures, including performance on cognitive tasks and answers to questionnaires, from the connections of each brain region (Figure 1a-c). In the course of this exercise, we are able to determine whether the functional forms of the relations between brain connectivity and observable subject measures are sufficiently complex so as to require deep neural networks. Moreover, instead of assuming that a single function exists between connectivity and behavior, we carve the brain at its joints with deep neural networks, generating a different function between each region's connectivity and each subject measure. These predictions can then by averaged, and we can test whether that agglomeration increases our prediction accuracy. Finally, deep neural networks are naturally modeled as multislice networks, as each layer of the network is only connected to itself, forming a "slice" of the deep neural network. We can thus optimize a multislice modularity quality index ($Q$) to assess the community structure of the trained deep neural networks (Figure 1a)[40]. A high $Q$ value indicates that the deep neural network is decomposable into communities with dense connectivity within each community and sparse connectivity between communities, indicating mostly independent paths through the deep neural network from the feature input layer to the prediction node. By building and structurally characterizing multislice network models[8,41] of the trained deep neural networks in the context of which brain region the model is capturing, and how accurate the model is, we are able shine light into the black box of deep neural networks to understand how and why they perform as they do.

We hypothesized that, while the function between connectivity and behavior for many regions and systems can be captured by linear regression, more complex functions that exist at the regional level require deep neural networks. We expect these more complex functions to exist at the brain's connector hubs[12], given their diverse connectivity across the brain's communities[2,3], their cortical expansion from macaques to humans[42–44], and their involvement in many different and more complex cognitive processes[4]. We quantify connector hubs on a continuum by the participation coefficient ($PC$)[2–4]. More generally, we reasoned that all regions are unlikely to share the same function between brain connectivity and behavior. Thus, we hypothesized that the best predictions across subject measures would be made while using a different deep neural network for each region, generating unique functions linking regional connectivity to behavior, and then averaging those predictions together. Finally, we hypothesized that the architectures of deep neural networks would have to be more modular while modeling connector hubs than while modeling non-connector hubs, in order to disentangle the different non-linear components of the function between connectivity and behavior.

**Deep neural networks outperform linear regression when using regional connectivity**. We first sought to measure the performance of linear and deep neural network models to predict subject measures from regional and system level connectivity. For each subject measure in the Human Connectome Project ($n$=52) and for each region in the brain ($n$=400[11]), we used that region's functional connections to predict that subject measure with deep neural networks and separately with linear regression (Figure 1a-c, Methods). We employed 5-fold cross-validation such that each prediction is generated by applying the trained model to the connectivity of untrained subjects (Figure 1b, Methods). We considered 100 distinct architecture for the deep neural networks, ranging from 1 to 10 layers and from 10 to 400 neurons (Figure 1c, Methods). We assessed prediction accuracy by calculating the Pearson $r$ correlation coefficient between the observed and predicted subject measures. To move beyond regional information, we also considered the 17 "Yeo" cognitive systems that were defined previously[11]; specifically, we used each system's connectivity to other systems as the 16 input features to the prediction algorithm. This approach allowed us to determine whether deep neural networks outperform linear regression only when the granularity of the connectivity feature space is high. In sum, for every subject measure (Figure 1d), we quantify how well we can predict that subject measure across 100 deep neural network architectures and separately with linear regression, both using regional connectivity and system connectivity.

Using system connectivity, we observed no significant differences between the prediction accuracy of deep neural networks and linear regression (Figure 2a). However, using regional connectivity, deep neural networks outperformed linear regression by a significant margin (Figure 2b). Next, we investigated which deep neural network predicted subject measures the most accurately, as well as which deep neural network architectures best captured the function between the connectivity of connector hubs and subject measures (Figure 2c,d). When considering either system or regional connectivity, deep neural network architectures with more layers and more neurons more accurately captured the connectivity of connector hubs. Notably, this result was less pronounced for system connectivity, where very small (1 layer, 10 neurons) deep neural network architectures produced accurate predictions for systems with many connector hubs (Figure 2c,d).

**Deep neural networks fit complex connector hubs**. Next, we sought to decipher where in the brain deep neural networks are more accurate than linear regression, and vice versa, relative to each model's mean prediction accuracy. We hypothesized that connector hub regions, given their diverse connectivity across the brain's communities, would require deep neural networks to discover the function between their connectivity and behavior. We thus measured each region's mean prediction accuracy across deep neural network architectures and subject measures, and we then $z$-scored those accuracy values. We also $z$-scored each region's mean prediction accuracy across subject measures for the linear regression models. Our choice to $z$-score was motivated by the goal to obtain a relative measure of where in the brain each model is least versus most accurate.

We found that regions with higher $z$-scored prediction accuracy for linear regression tended to be located in visual, sensory, and motor cortex (Figure 2e), while regions with higher $z$-scored prediction accuracy for deep neural networks tended to be connector hubs located in fronto-parietal cortex (Figure 2f,g). To further determine whether deep neural networks are most accurate at connector hub regions, we calculated the Pearson $r$ correlation coefficient between each region's prediction accuracy and participation coefficient, for each subject measure. We found that the $r$ values were significantly larger for deep neural networks than for linear regression (Figure 2h). Moreover, we observed a positive and significant Pearson $r$ correlation coefficient between each region's mean prediction accuracy across subject measures for deep neural networks and each region's participation coefficient (Figure 2i). This relationship held true even after subtracting the mean prediction accuracy across subject measures for linear regression models (Figure 2j). Taken together, these results demonstrate that deep neural networks, in general and compared to linear regression, generate the most accurate predictions of subject measures from connector hub connectivity.

**Connector hubs are modeled by modular deep neural networks**. Next, we sought to understand how the architecture of a deep neural network can support accurate predictions. In particular, we were interested in discovering why deep neural networks are able to model connector hub regions so accurately. We modeled the deep neural network as a multislice network in which each hidden layer's neurons and connections between them comprise a slice and each neuron is connected to itself across slices. Moreover, in order to isolate phenomena that are driven by training the deep neural networks on brain connectivity and behavior from deep neural networks in general, for each subject measure, we shuffled the values and trained a deep neural network to predict random subject measure values from brain connectivity. Next, we assess the multislice network's architecture by maximizing a multislice modularity quality index, $Q$, to identify communities of neurons that exists across the slices[40] (Methods, Figure 1a,c). Here, a community is a collection of neurons that are tightly connected to each other as one traverses the layers of the deep neural network. An intuitive notion of this type of community is a path or stream involving multiple neurons through the layers of the deep neural network, as these neurons remain tightly interconnected across layers. A set of communities would therefore indicate parallel streams of neurons across layers that allow for simultaneous processing of distinct functions.

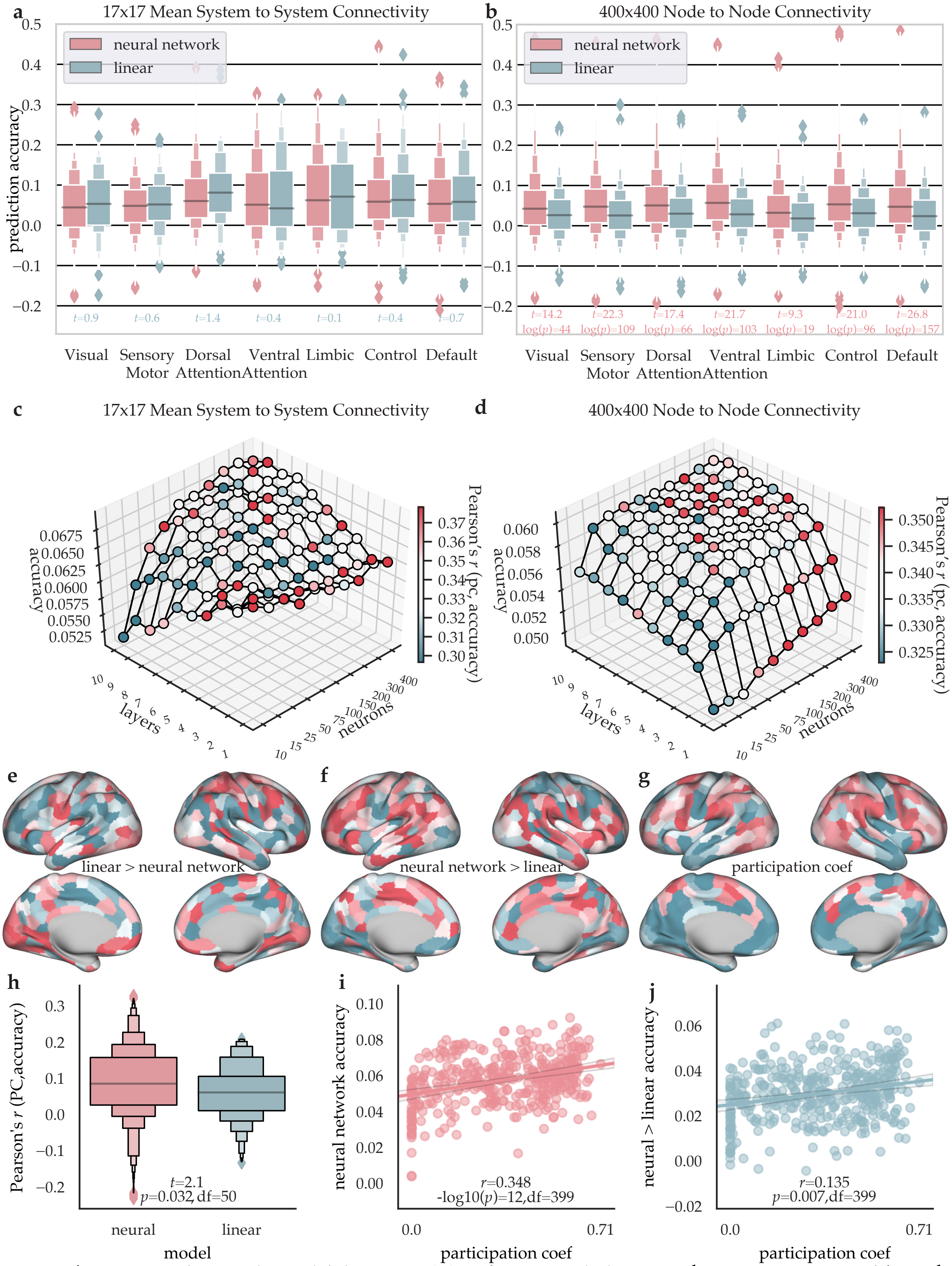

Figure 2 | **Deep neural networks model the connectivity of connector hubs**. For each region or system and for each subject measure, we calculate prediction accuracy with the Pearson $r$ correlation coefficient between the observed subject

measures and the predicted subject measures. **a**, Prediction accuracy value distributions using each system's connectivity to other communities, shown grouped by large scale Yeo system ($n$=7), for both deep neural networks and linear models. We used students independent $t$-tests with Bonferroni corrected $p$-values to test for differences between linear and deep neural network models in terms of prediction accuracy at each system. While $t$-values were positive for all linear models, none were significant ($t$-values < 1.4, $p$>0.05, 5250<df>21006). **b**, Prediction accuracy value distributions using each region's connectivity to all other regions, shown grouped by large scale system. For regions in every system, prediction accuracies were significantly higher for deep neural networks than for linear models ($t$-tests,126046<df>498938). **c,d** Neural network architectures are plotted on the x-axis (neurons) and y-axis (layers). The mean prediction accuracy is shown on the z-axis. Each region is colored according to the Pearson $r$ correlation coefficient between each system's (c) or region's (d) mean prediction accuracy across subject measures and the system's or region's participation coefficient. Note that for panel (c), we used the mean participation coefficient of regions in the system. In panel **c**, results are shown from deep neural networks that model each system's connectivity to the other systems; in panel **d**, the deep neural networks model each region's connectivity to the other regions. The mean prediction accuracy is taken across systems (c) or regions (d) and then across measures. From panel (d), we observe that wider deep neural network architectures best model connector hub connectivity, as deep neural networks with 100 or more neurons had significantly higher Pearson $r$ correlation coefficients between each region's accuracy and each region's participation coefficient ($t$=6.82,-log10($p$)=10,df=98). We see this effect visually by the high accuracy at regions with high participation coefficient (deeper red along the color bar). **e**, We calculated the mean prediction accuracy values across subject measures and deep neural network architectures for each region as well as the mean prediction accuracy across subject measures at each region for the linear models. We then z-scored these values and subtracted the values obtained from the deep neural network values from those obtained from the linear model values. **f**, As in panel (e), except that we subtracted the values obtained from the linear models from those obtained from the deep neural networks. **g**, The participation coefficient for each region is calculated on the mean (across subjects) connectivity matrix (see Methods). **h**, We compared the distribution of Pearson $r$ correlation coefficients between each region's prediction accuracy and participation coefficient for each subject measure and for each deep neural network architecture ($n$=5200) to the distribution of $r$ values between each region's linear regression prediction accuracy and participation coefficient across subject measures (n=52). On average, these $r$ values are significantly larger for the deep neural networks. **i**, The Pearson correlation between each region's mean accuracy across subject measures and deep neural network architectures and each region's participation coefficient. **j**, We calculated the Pearson $r$ correlation coefficient between values from panel (f), which assess where in the brain deep neural networks are more accurate than linear models, and each region's participation coefficient as shown in panel (g). This result can be visualized by comparing the values in panels (f) and (g).

We found that deeper (more layers) and wider (more neurons) neural networks exhibited higher Q values (Figure 3a).This observation was true of deep neural networks in general, as both deep neural networks trained on the real and shuffled subject measures exhibited this effect. Moreover, the Q value for each architecture was highly similar, regardless of whether the deep neural network was trained on original or shuffled subject measures (Pearson's r=0.985,-log10(p)>100,df=48; Extended Data Figure 3a). Only when using the real (not shuffled) data, the same deep and wide architectures that exhibited high Q exhibited high prediction accuracy (Figure 3b,d; Extended Data Figure 3b). Interestingly, the most accurate architectures for connector hubs were not necessarily deep, but wide (Figure 3c); again, this effect was not present when the deep neural networks were trained on shuffled subject measures (Extended Data Figure 3c). To investigate this architectural specificity further, we calculated a polynomial regression between each network's Q and prediction accuracy (Figure 3d). In order to know if modeling a connector hub well drives the deep neural network architecture to be extra modular, we analyzed the residuals from the exponential fit—positive residuals mean that the network is more modular than it has to be, in general across brain regions, in order to be accurate. We found that the residuals from this function were significantly higher for architectures that modeled the connectivity of connector hubs well (Figure 3d,e). Critically, this effect was not present when the deep neural networks were trained on shuffled subject measures (Extended Data Figure 3d,e). Moreover, across subject measures, the modular deep neural network architectures were the most accurate if they also modeled the connectivity of connector hubs well (Figure 3f). Again, this effect was not present when the networks were trained on shuffled subject measures (Extended Data Figure 3f). Thus, it appears that modeling the connectivity of connector hubs drives deep neural networks to be extra-modular.

Finally, connector hubs, because they are widely connected to many communities across the brain, potentially contain information about the entire brain's connectivity[2–4,18]. If true, we reasoned that a deep neural network architecture should be able to predict a subject measure quite well from a single region, so long as that region is a connector hub[2–4,18]. Consistent with our expectation, we found that the deep neural network architectures that were able to most accurately model the connectivity of connector hubs were also able to outperform the combined prediction accuracy measure (Figure 3g,h). These architectures were all wide (Figure 3g). This effect was not present when the deep neural networks were trained on shuffled subject measures (Extended Data Figure 3g,h). Thus, a single region can outperform whole brain predictions, so long as the deep neural network architecture is wide (many neurons) and the region is a connector hub.

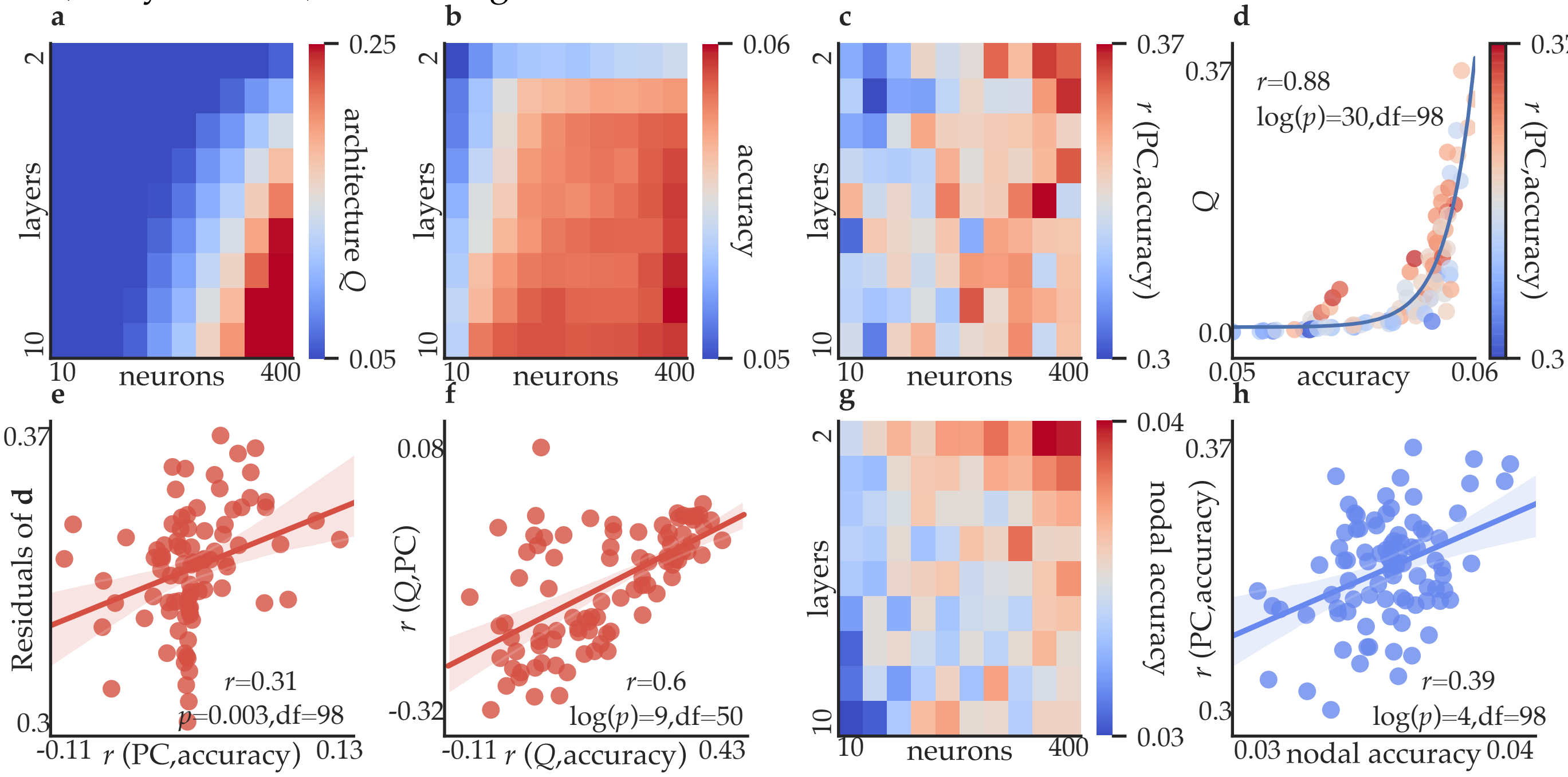

Figure 3 | **Modular deep neural network architectures are the most accurate and model connector hubs**. **a**, We build a multislice network model of each deep neural network, where each set of neurons and connections within a hidden layer is a slice, and where connections are placed between a neuron and itself across slices. We then maximize a modularity quality index ($Q$) to assess the structure of this multislice network model. Deep neural network architectures with more layers and neurons exhibit a higher $Q$. **b**, The mean accuracy across subject measures for each architecture is shown. Deep neural network architectures with more layers and more neurons exhibit higher accuracy. **c**, For each subject measure and architecture, we calculated the Pearson correlation between each region's accuracy and each region's participation coefficient. The mean across all subject measures is shown. **d**, An exponential fit of the relationship between each architecture's $Q$ (matrix from panel a) and each architecture's accuracy (matrix from panel b), with values colored according to the Pearson correlation coefficient between each region's prediction accuracy and each region's participation coefficient (matrix from panel c). **e**, The Pearson correlation between the residuals from panel (d) and the extent to which each architecture models connector hubs more accurately than other regions (matrix from panel c). Thus, deep neural network architectures are more modular than average while modeling connector hub connectivity. **f**, The Pearson correlation between (i) the extent to which a subject measure is modeled by modular architectures of connector hubs (y-axis; Pearson's $r$ between nodal participation coefficient and architecture $Q$ for that region), and (ii) the extent to which a subject measure is best modeled by architectures that are modular (x-axis; Pearson's $r$ between nodal accuracy and architecture $Q$ for that region). Thus, subject measures that are best modeled by modular architectures tend to rely on modular architectures of connector hubs for their model accuracy. **g**, Prediction accuracy can either be made by taking the mean accuracy across regions for each subject measure, or by calculating the mean prediction across regions for each subject, creating a combined prediction of the subject measure. The difference between the combined prediction and the most accurate single region prediction, averaged across subject measures, is shown. **h**, The Pearson correlation between the values from panel (g) and the values from panel (c), demonstrating that architectures modeling a single region are able to outperform the combined accuracy if they are able to model connector hubs well.

**Combined deep neural networks best predict behavior**. If deep neural networks are carving the brain at its joints, generating the best unique function for each region, then averaging the predictions across regions to generate a single prediction of the subject measure across individuals should be highly accurate. To test this expectation, we generated a combined prediction for each subject measure and for each deep neural network architecture, where the mean prediction across regions for that subject measure is calculated; we calculate the mean, across regions, of the 400 (regions (or 17 systems)) x 607 (subjects) prediction array, generating a 1x607 prediction array. This averaging simultaneously leverages the unique functions at each region and reduces prediction noise from overfitting. In contrast to deep neural networks, we expect a simple linear model to, on average, generate the best predictions from individual systems, because averaging regional connectivity down to system level connectivity removes any complex non-linear functions between connectivity and behavior. To test this expectation, we compared the distributions of prediction accuracies for system and regional predictions across linear and deep neural networks, both from individual systems and regions as well as while averaging the predictions across systems or regions. Recapitulating what we observed above (Figure 2a), we found that the deep neural networks outperform the linear models for regional connectivity (Figure 4a). In contrast, the linear models for system connectivity outperform the deep neural network models for regional connectivity (Figure 4a). Finally, the system models outperform the regional models, both for linear and deep neural networks (Figure 4a). In sum, when making a prediction from a single region or system, the most accurate predictions come from models of system connectivity, with linear models being numerically but not significantly more accurate than deep neural networks.

Notably, whereas the mean accuracy is highest for models of system connectivity, deep neural networks for regional connectivity still generated the highest accuracies we observed, with many $r$ values above 0.40 (Figure 4b). In fact, the best prediction accuracies for each subject measure for deep neural networks of regional connectivity were significantly higher than the linear models' top prediction accuracies for each subject measure (student's independent $t$=6.60, -log10($p$)=9, $df$=102). Thus, deep neural networks trained on regional connectivity are able to find the strongest relationships between the brain and behavior, particularly at connector hubs, even though they do not perform optimally, on average, across all regions in comparison to linear models of system connectivity (Figure 4a). As described above, however, we can make a combined average prediction that leverages both this optimality at connector hubs and the ability of deep neural networks to fit a function to every brain region. When making combined average predictions, utilizing information across all regions or systems, deep neural networks applied to regional connectivity perform better than any other model tested here (Figure 4b). For each subject measure, we visualized the distributions of the prediction accuracies, comparing the combined deep neural network prediction accuracies to the most accurate deep neural network architectures and linear models for a given subject measure from a single region. In general, we found that no model could outperform the combined predictions from deep neural networks trained on regional connectivity. Thus, the most accurate model of brain connectivity and subject measures appears to actually be a large collection of deep neural networks, one for each brain region.

In sum, deep neural networks trained on regional connectivity carve the brain at its joints, generating complex, but meaningful, functions that are unique to each region. While deep neural networks best model the connectivity of connector hubs, they can also model other regional connectivity well, and, by combining all of these predictions, we obtain the most accurate predictions of human behavior. Critically, neither of these characteristics exist for linear models.

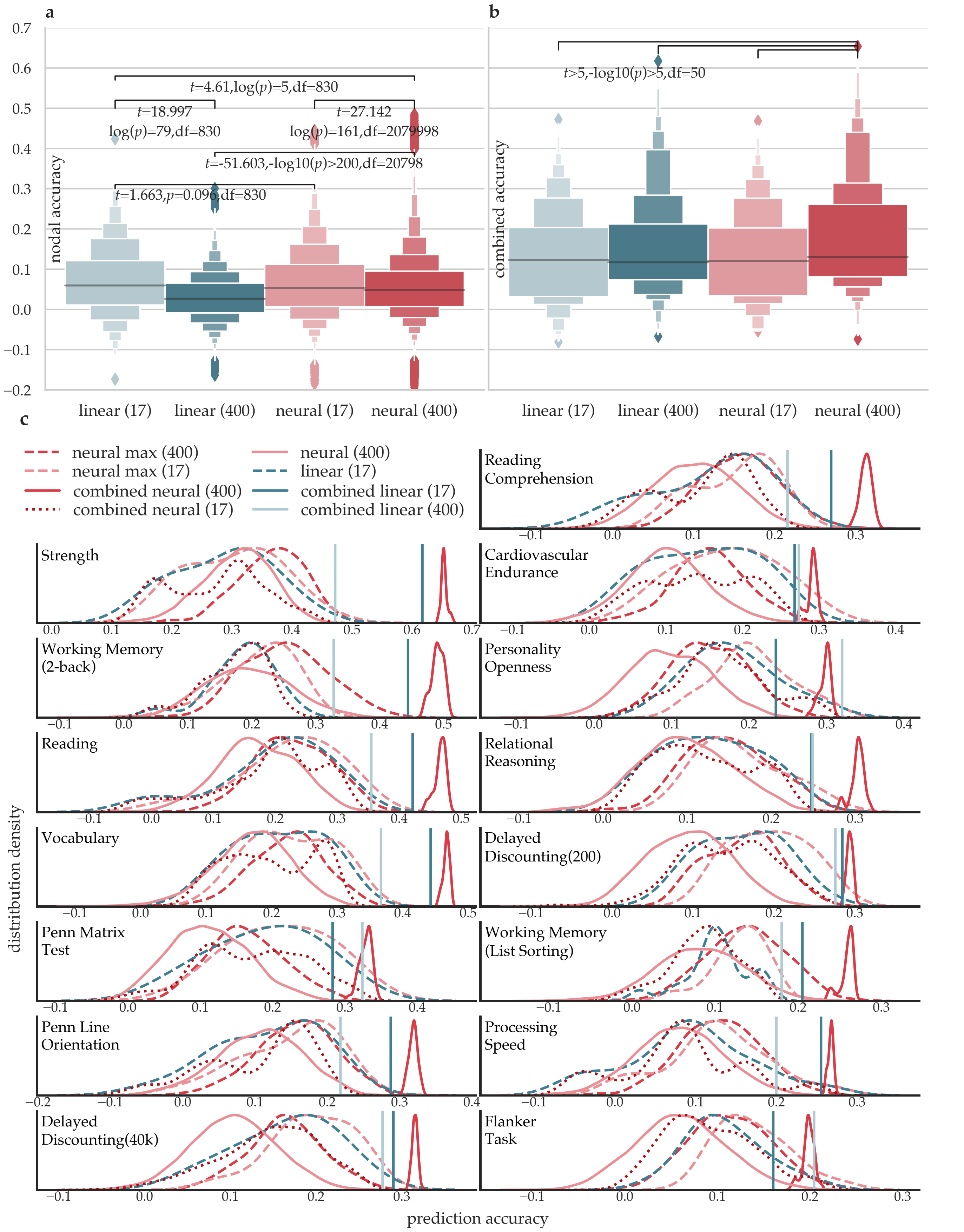

Figure 4 | **Combining the predictions of deep neural networks across regions yields the most accurate predictions**. **a**, Distributions of the prediction accuracy for system and regional predictions. Whereas there is no significant difference between the prediction accuracy of linear models and deep neural networks when trained on system connectivity, and the deep neural networks outperform the linear models when trained on regional connectivity, the linear models of system

connectivity outperform the deep neural network models trained on regional connectivity. **b**, Each model, linear or deep neural network, makes a prediction for each subject for a given subject measure. Thus, for the models trained on system connectivity, there are 17 predictions for each subject, and, for the model trained on regional connectivity, there are 400 predictions for each subject. The mean prediction across the 17 systems or 400 regions is the combined prediction. The distribution of these prediction accuracies is shown across tasks. When making combined predictions, utilizing information across all regions, deep neural networks trained on regional connectivity perform the best. **c**, Distributions of the prediction type accuracies across all measures and then for the most accurately predicted measures. "Neural max" shows the most accurate deep neural network architecture for each region for that subject measure. In general, deep neural networks combined at the regional level make the most accurate predictions.

**Deep neural networks versus ridge regression with nested cross-validation**

Our primary goal here, particularly as neuroscientists, was to decipher whether non-linear functions exist between brain connectivity and behavior, making deep neural networks principled for use. We thus analyzed how deep neural networks, in general across architectures, compared to linear models. However, even though the deep neural networks appear to be the most *principled* approach, our next goal is to maximize the prediction *accuracy* of the models we tested. Here, the best deep neural network architecture can be selected for prediction. Moreover, instead of linear regression, we can use ridge regression, which is simply linear regression with L2 regularization. In our previous results, we used linear regression, which fits linear functions to every connection and takes the coefficients at face value. This process allowed us to demonstrate that there are non-linear functions which the linear regression model misses or misinterprets, but the deep neural networks, in general across architectures, can accurately leverage. Here, we go one step further to ask a slightly different, more applied machine learning, question: Does modeling linear functions, with regularization via ridge regression to shrink the coefficients, lead to a more accurate prediction than using the most optimal deep neural network architecture? It is plausible (but not yet proven) that deep neural networks are the most principled, which we showed above, *and* the most accurate.

First, for each brain region and task, we used nested cross-validation to find the optimal L2 regularization parameter (a) in ridge regression. Thus, within each fold, on the training set, we find the □□that maximizes prediction across folds within that training set, and then that □□is used to make the prediction on the test set. We found that the *average* deep neural network (i.e., no nested cross-validation to find the best architecture) typically outperforms the *best* ridge regression (as in the model with the most accurate L2 regularization parameter found via cross-validation within each training fold) across the 52 HCP behaviors when making combined predictions, but not when prediction from single nodes (Extended Data Figure 1; $t$=3.45,-log10($p$)>5).

Next, we sought to find the best deep neural network architecture and ridge regression model using nested cross-validation for model selection for both models. Thus, here, we find a single model for each task and node, where the deep neural network model has the most accurate architecture and the ridge regression model has the most accurate L2 regularization parameter, both found in the training folds via cross-validation. Here, we found similar results again, in that ridge regression outperforms deep neural networks at the regional level, but not when making combined predictions (Extended Data Figure 2).

Finally, we sought to test deep neural networks and ridge regression against each other in the UK BioBank (UKB), analyzing the fluid intelligence performance. Brain connectivity in the UKB was generated using the same preprocessing as the HCP (see Methods). Using the HCP, we find the deep neural network architecture model and ridge regression regularization model that perform best in the HCP working-memory task. We chose working memory, as it is a behavior similar to fluid intelligence and was also well predicted by our models. Thus, for each brain region, we find the alpha (ridge regression) or network architecture (deep neural network) that best predicts working memory in the HCP. We will then apply these models, for example, for a visual cortex region, a deep neural network with 400 neurons and 10 layers, and a ridge regression model with an alpha regularization parameter of 10, to that region's brain connectivity and the fluid intelligence measure in the UKB. Again, we find that ridge regression is much more accurate at the regional level, but the two methods are essentially equal when making combined predictions from whole brain connectivity (Figure 5).

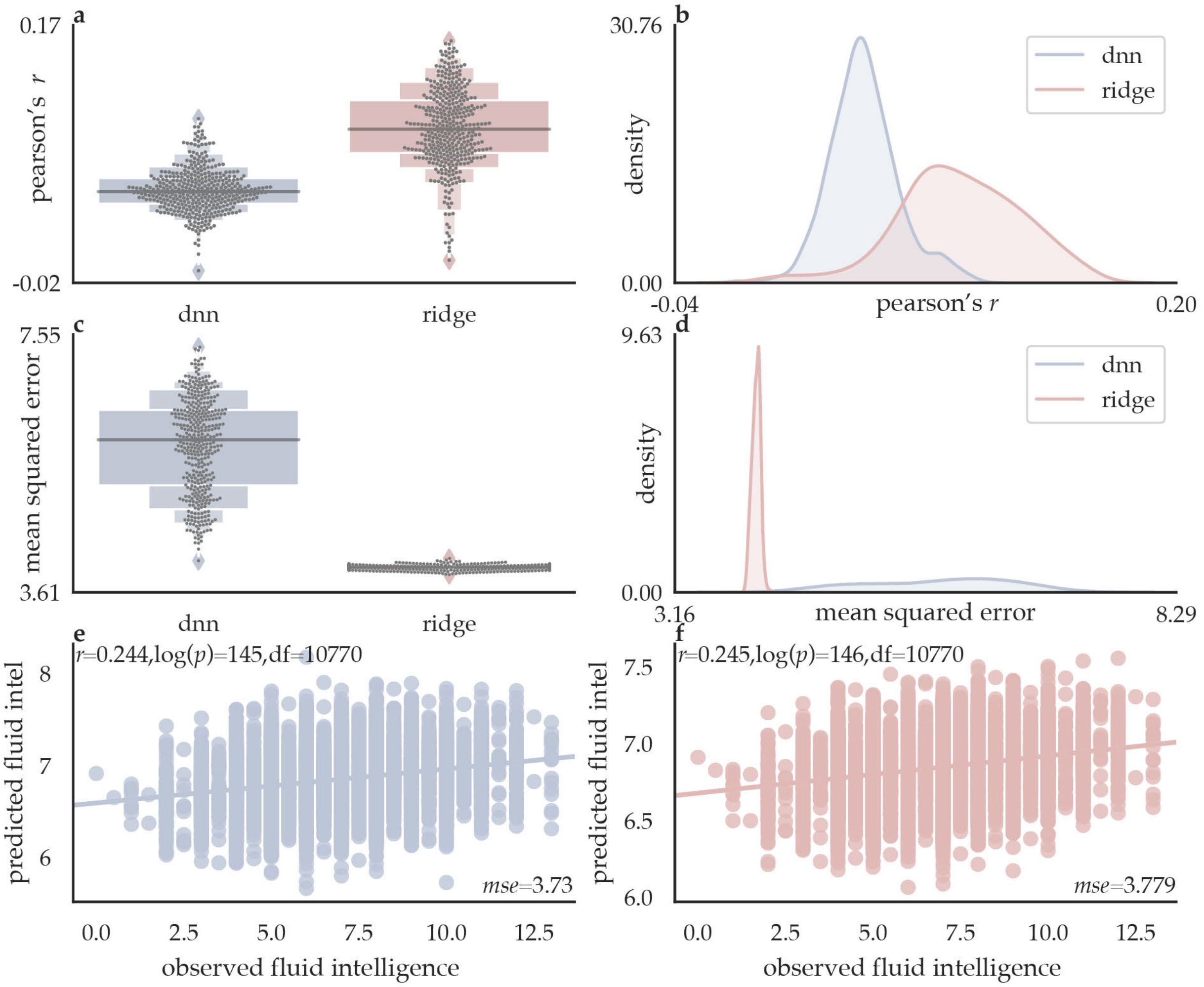

Figure 5 | **Deep neural networks and ridge regression models of fluid intelligence in the UK BioBank**. In the UKB, for each brain region, the deep neural network and ridge regression model that optimally predicted working memory performance in the HCP are used to model brain connectivity and fluid intelligence in the UKB. **a-d**, The distribution of Pearson *r* (**a**,**b**) and mean squared error (**c**,**d**) accuracy values across brain regions. **e**,**f**, The Pearson *r* correlation between the observed and predicted fluid intelligence scores in the UKB.

**Discussion**. We discovered that there are complex relationships between brain connectivity and behavior that are best captured by deep neural networks and we can use network science to interpret the architectural properties of the most accurate deep neural networks. Deep neural networks, particularly ones with many neurons, become modular in order to most accurately model the connectivity of connector hubs. Connector hubs have been demonstrated to integrate information across the brain's systems, as well as coordinate connectivity between the brain's systems, allowing for a balance of modular and integrative processing[2–4,7,16–18,23,45–47]. The connectivity of these regions thus likely integrates behaviorally relevant information from across the entire brain. Using multislice network models to measure the modularity of trained deep neural networks, we find that wide and modular deep neural networks decompose the complex connectivity of connector hubs optimally, likely by creating mostly independent communities or parallel paths through the network to best learn the nonlinear relations between connector hub connectivity and behavior.

Past research has shown that modular biological networks emerge when varying the training goals of the system[48] and modular neural networks exhibit faster information spreading and more precise memory[49]. Here, we add that, in the context of brain connectivity and behavior, deep neural networks perform better when their architecture is modular, particularly for the brain's connector hubs, likely because these regions have diverse connectivity to the brain' communities. The deep neural network architectures are able to decompose each connector hub's connectivity to individual communities while also modeling the relevant dependencies between the way in which the connector hub is connected to each community. Potentially because of this modeling ability, predictions from connector hubs are often better than predictions utilizing the entire brain's connectivity. Despite highly accurate predictions made from connector hubs in certain deep neural network architectures, we found that, in general across deep neural network architectures, the most accurate predictions of behavior come from combining the predictions of all regions.

Finally, when comparing deep neural networks to ridge regression, we find that ridge regression is more accurate when modeling a single region's connectivity, but deep neural networks are more or equally accurate, depending on the measurement, when modeling whole brain connectivity. Deep neural networks are able to model any function; while this likely is noisier than ridge regression at the regional level, it becomes advantageous at the whole brain connectivity level, as the deep neural networks can model all the unique functions of the brain, and noise is reduced when averaging predictions across the brain.

In sum, instead of requiring a single function to explain the relation between brain connectivity and behavior, deep neural networks are able to carve the brain along multiple unique joints, from simple linear ones to complex nonlinear ones, that exist between the brain and behavior. Critically, our results motivate future studies to determine whether deep neural networks could better decompose the complex functions that exist between brain connectivity and mental illness[9,10,37,26,50–53]. While ridge regression and deep neural networks are oftentimes similar in accuracy, deep neural networks can certainly be preferred given a particular scientific question. Here, we demonstrate this by using deep neural networks to understand the connectivity and function of connector hubs.

**Methods**

*Subject Measures*. We chose to analyze a large and diverse set of measures in the Human Connectome Project to ensure that our conclusions were valid across distinct assessments of cognition and behavior. We included all measures from the Alertness, Cognition, Emotion, Personality, and Sensory categories. In addition, we used several in-scanner measures of behavioral performance on cognitive tasks. Extended Data File 1 includes the full list of measures and our more natural language names for them.

*Human Brain Connectivity*

Human Connectome Project. The Resting-state, Working Memory, Language and Math, Relational, Social Cognition, Emotion, Gambling, and Motor fMRI scans from the S1200 Human Connectome Project release[54] were analyzed. Because our analyses included many different subject measures, we chose to use all scans in order to sample the brain's functional connectivity in a diverse range of cognitive states. Additionally, this choice likely increases the accuracy of our measurement of each functional connection, given the increase in the length of the time series. Brains were normalized to fslr32k via the MSM-AII registration. CompCor, with five principal components from the ventricles and white matter masks, was used to regress out nuisance signals from the time series. In addition, the 12 detrended motion estimates provided by the Human Connectome Project were regressed out from the regional time series. The mean global signal was removed and then time series were band-pass filtered from 0.009 to 0.08 Hz[52-53]. Finally, frames with greater than 0.2 mm frame-wise displacement or a derivative root mean square (DVARS) above 75 were removed as outliers. Segments of less than five uncensored time points were also removed. Sessions composed of greater than 50 percent outlier frames were not further analyzed. The processing pipeline used here has previously been suggested to be ideal for removing false relations between connectivity and behavior[49]. We then only analyzed data from subjects that had all scans remaining after scrubbing, leaving 607 subjects.

UK Biobank. UKB analyses were conducted using the UKB Resource under application 22875. Image acquisition for the UK Biobank dataset has been detailed elsewhere[55,56]. Briefly, structural and functional images were collected across three imaging sites with identical scanners (3T Siemens Skyra, software VD13) and protocols. Prior to distribution, an automated quality control tool examined images from each modality to verify correct dimensionalityo[55]. For consideration into the current study, subjects must supply usable T1w, T2w, and rfMRI timeseries images. Structural T1w and T2w images were processed using the *PreFreeSurfer* and *FreeSurfer* stages of the Human Connectome Project's (HCP) minimal preprocessing pipeline[57]. The *Prefreesurfer* stage corrects each image for gradient distortions and then co-registers the two structural images. During the *FreeSurfer* stage an HCP-specific distribution of Freesurfer is used (5.3.0-HCP) to generate cortical surfaces, with the T2w image aiding in the pial surface generation. We used the preprocessed rfMRI timeseries data distributed through the UK Biobank[55]. These data were motion corrected, intensity-normalized, highpass filtered with a sigma of 50 seconds, and corrected for gradient field distortions. Then, structured artifacts were removed with ICA-FIX using a classifier that was hand-trained using rfMRI data from 40 UK Biobank subjects[58,59]. The rfMRI timeseries was then co-registered to the structural images produced from Freesurfer. We applied a bandpass filter (0.009 – 0.08 hz) and global signal regression, and then projected the rfMRI timeseries to the cortical surface.

For all subjects, we parcellated the brain into 400 regions[11] on the fslr32k cortical surface and calculated the Pearson correlation coefficient between all pairs of regions to generate a 400x400 matrix for each scan; to each element we applied Fisher's r-to-z transformation (numpy.arctanh). For the HCP, these 18 matrices per subject (four for the resting-state, and two for each task) were then averaged for each subject into a final 400x400 matrix. To calculate the participation coefficient, we took the mean connectivity matrix across all subjects and scans. Across 50 linear steps from thresholds of $90^{th}$ to $95^{th}$ percentiles (i.e., graph densities from 0.1 to 0.05), we calculated the weighted participation coefficient, where the canonical 17 systems were used as communities[11]. The mean participation coefficient across thresholds was then calculated and used for all analyses.

*Machine Learning*
We used python's sklearn library to train deep neural networks, linear models, and support vector machines. For the deep neural networks, we used: sklearn.neural_network.**MLPRegressor** with stock parameters, except for using " solver='lbfgs' ", as, for small datasets, 'lbfgs' can converge faster and perform better (see sklearn documentation). We then trained networks in all combinations of 1-10 layers in steps of 1 and 10, and 10, 15, 25, 50, 75, 100, 150, 200, 300, and 400 neurons. Linear regression was executed with stock parameters as: sklearn.linear_model.**LinearRegression**. For nested cross-validation, Ridge Regression was ran as: sklearn.linear_model.**RidgeCV**, with alphas set between 1e-10 to 100, and every architecture of deep neural network was tested within each fold. For every subject measure, we used 5-folds cross validation (sklearn.model_selection .KFold(5, shuffle=True) to generate 5 folds, and then we ran each model—deep neural network, linear, and ridge regression—on the same division of subjects into 5 folds, as prediction accuracy can vary if the division of subjects into folds differs. We then measured prediction accuracy as the Pearson *r* correlation coefficient between the predicted and observed subject measure. In line with previous and similar studies that compare machine learning methods, in order to maximize the number of subjects, twins were included[2,34]. We note that, critically, we used the same division of subjects into folds for all models, ensuring that any potential performance advantage due to twins in the data was equally present in all prediction accuracy measures. This approach also ensures that differences across predictions are not due to differences in folds.

*Computing*
Deep neural networks are notoriously expensive to train. Here, we levered the CUBIC cluster at the University of Pennsylvania, which consists of 168 compute nodes, providing a total of 4804 CPUs. Each deep neural network model used, at most, 4GB of RAM to run. On this cluster, running all regions ($n$=400), all subject measures ($n$=52) and all architectures ($n$=100), takes approximately one week. Including the multislice modularity analyses, linear models, and system level predictions, the analyses here take approximately two weeks on our cluster, or roughly 1,600,000 hours of compute time.

*Multislice Modularity*
A multislice network is composed of individual networks coupled through links that connect each node in one network slice to itself in other slices[41]. One example of a common multislice network is a network that changes through time, where each point in time is represented as a "slice". A deep neural network can also naturally be seen as a multislice network, where each hidden layer is a slice, and weights link nodes to one another, forming a neuron-by-neuron connectivity matrix. Moreover, each neuron is connected to itself in the previous and next layer. In order to model the strongest edge weights in the deep neural network, the matrix is thresholded to only retain weights above the 90$^{th}$ percentile, which is the median of our typical set of thresholds[2–4], while ensuring that every node has at least one edge by calculating the maximum spanning tree[2]. Edge weights are not binarized. In contrast to when we predict subject measures with 5-fold cross validation, here the deep neural network is trained on the full set of subjects in order to generate a single deep neural network that models the function between brain connectivity and the given subject measure for all subjects. This is also necessary, as we sought to analyze a single deep neural network for each region and subject measure. For multislice modularity[40], where we detect communities across slices, each neuron is linked to itself across slices by the $\Omega$ parameter, which is set to 0.5. We chose this value because the mean weight in the multislice network is roughly 1, so a $\Omega$ of 0.5 maintains the node's connectivity to itself throughout layers while allowing the node to change communities. Finally, $\gamma$ is a resolution parameter that determines the resolution at which communities are detected, which is set to 2. We chose this value to ensure that no neural network produced less than three communities, nor was the network partitioned such that each node is its own community (for examples, see Figure 1c).

Although each parameter value was reasonably chosen to result in interpretable partitions of deep neural network nodes into communities, ideally, all three parameter choices, as well as a range of graph thresholds, are tested across a range of similarly reasonable choices. However, these parameters had to be chosen *a priori*, because, even with our current computing power, this is simply not feasible.

*Code and Data Availability*
Subject matrices and full analysis code can be found at: github.com/mb3152/deep_prediction.

Extended Data Figures

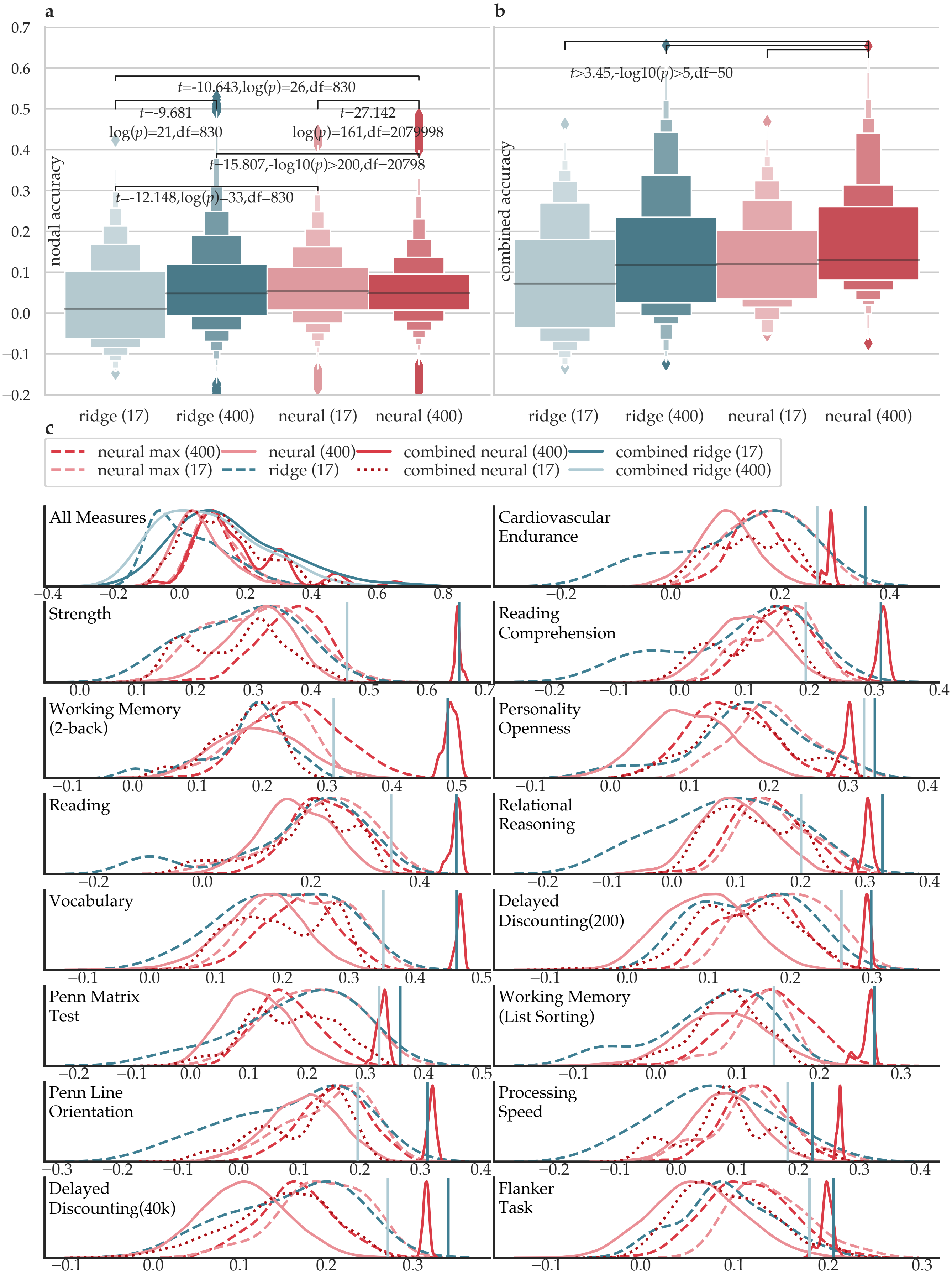

Extended Data Figure 1 | **Comparing deep neural networks and ridge regression**. **a**, The ridge regression models outperform the deep neural network models when trained on regional level connectivity, while the deep neural networks outperform the ridge regression models when trained on system level connectivity. **b**, Each model, whether a ridge regression or deep neural network model, makes a prediction for each subject for a given subject measure. Thus, for the models trained on system connectivity, there are 17 predictions for each subject, and, for the model trained on regional connectivity, there are 400 predictions for each subject. The mean prediction across the 17 systems or 400 regions is the combined prediction. The distribution of these prediction accuracies is shown across tasks. When making combined predictions, utilizing information across all regions, deep neural networks trained on regional connectivity perform the best. **c**, Distributions of the prediction type accuracies across all measures and then for the most accurately predicted measures. In general, deep neural networks combined at the regional level make the most accurate predictions.

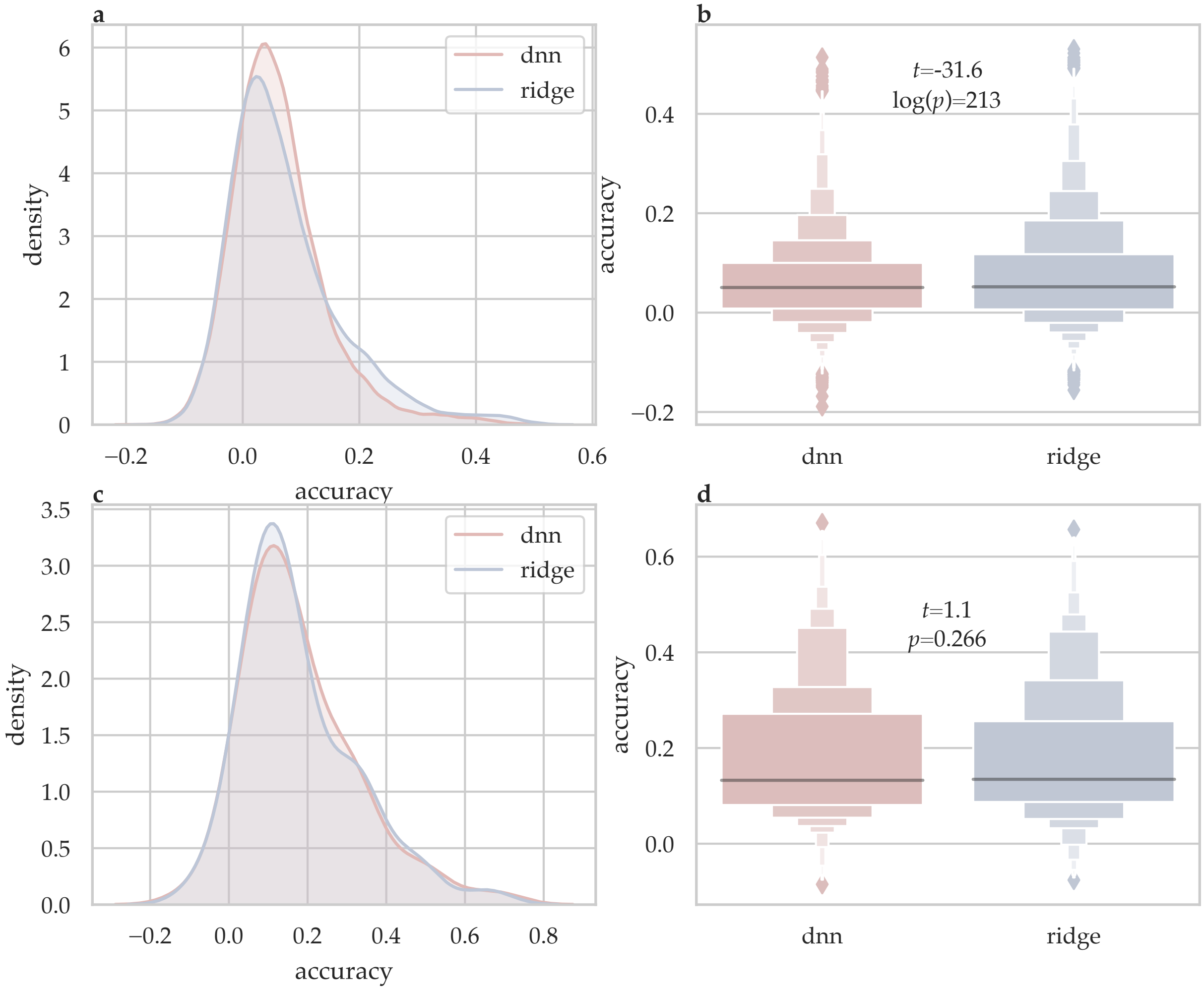

Extended Data Figure 2 | **Comparing deep neural networks and ridge regression with nested cross-validation**. For each task, the optimal deep neural network architecture and ridge regression alpha are found via nested cross-validation. **a**, the distribution in accuracy scores (Pearson *r*) for each node across all tasks. b, The values in (**a**) plotted with seaborn's "boxenplot"; the ridge regression models' accuracy, was, on average across all tasks and brain regions, significantly more accurate than the deep neural networks. For each task, the mean prediction across all brain regions can be made. **c**, the distribution in accuracy scores (Pearson *r*) for all tasks. **d**, The values in (**c**) plotted with seaborn's "boxenplot"; the deep neural networks' accuracy, was, on average across all tasks, more accurate than the the ridge regression models', but not a significant amount.

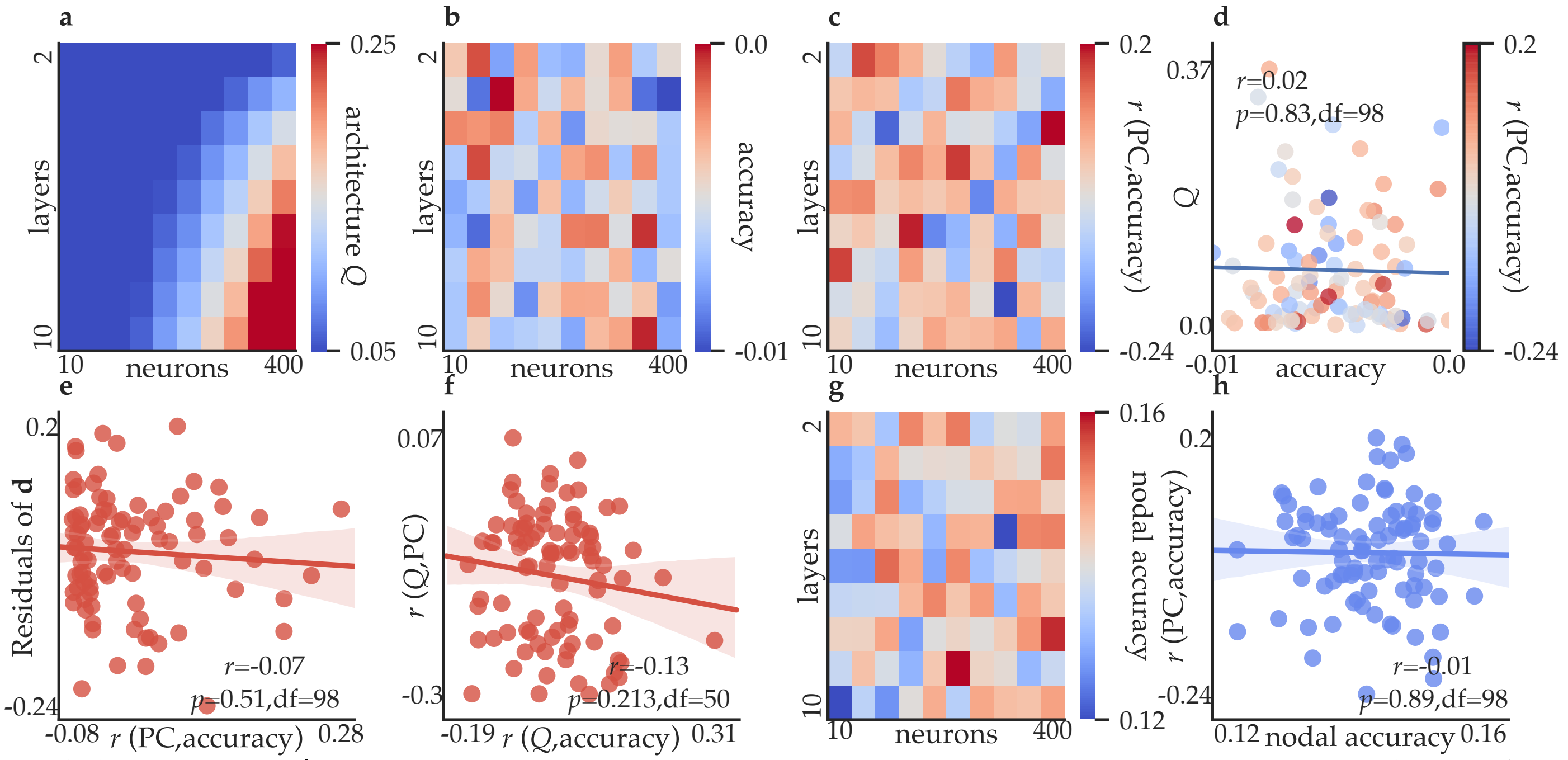

Extended Data Figure 3 | **Modular deep neural network architectures of random subject measures**. In order to isolate phenomena that are driven by training the deep neural networks on brain connectivity and behavior from deep neural networks in general, for each subject measure, we shuffled the values and trained a deep neural network to predict the random subject measure values from brain connectivity. **a**, We build a multislice network model of each deep neural network, where each set of nodes and connections within a hidden layer is a slice, and where connections are placed between a neuron and itself across slices. We then maximize a modularity quality index ($Q$) to assess the structure of this multislice network model. Deep neural network architectures with more layers and neurons exhibit a higher $Q$. **b**, The mean accuracy across subject measures for each architecture is shown. Unlike the case for unshuffled subject measures, accuracy is not dependent on the deep neural network's architecture (Figure 3). **c**, For each subject measure and architecture, we calculated the Pearson $r$ correlation between each region's accuracy and each region's participation coefficient. The mean across subject measures is shown; in contrast to the same analysis using unshuffled subject measures (Figure 3), no particular architecture is better at modeling connector hubs' connectivity if the subject measures are randomly shuffled. **d**, An exponential fit of the relationship between each architecture's $Q$ (matrix from panel a) and each architecture's accuracy (matrix from panel b), with values colored according to the Pearson correlation coefficient between each region's prediction accuracy and each region's participation coefficient (matrix from panel **c**). **e**, The Pearson correlation between the residuals from panel (d) and the extent to which each architecture models connector hubs more accurately than other regions (matrix from panel c); thus, deep neural network architectures are not more modular than they need to be in order to be accurate while modeling connector hub connectivity, which was the case with the unshuffled subject measures. **f**, The Pearson correlation between (i) the extent to which a subject measure is modeled by modular architectures of connector hubs (y-axis; Pearson's $r$ between nodal participation coefficient and architecture $Q$ for that region), and (ii) the extent to which a subject measure is best modeled by architectures that are modular (x-axis; Pearson's $r$ between nodal accuracy and architecture $Q$ for that region). In contrast to using unshuffled subject measures(Figure 3), shuffled subject measures that are best modeled by modular architectures do not tend to rely on connector hubs for their model accuracy. **g**, Prediction accuracy can either be made by taking the mean accuracy across regions for each subject measure, or by calculating the mean prediction across regions for each subject, creating a combined prediction of the subject measure. The difference between the combined prediction and the most accurate single region prediction, averaged across subject measures, is shown. **h**, The Pearson correlation between the values from panel (g) and the values from panel (c), demonstrating that architectures modeling a single region are not able to outperform the combined accuracy if they are able to model connector hubs well, which was the case when real subject measures were predicted(Figure 3).

## Citations

**Citation Diversity Statement** Recent work in neuroscience and other fields has identified a bias in citation practices such that papers from women and other minorities are under-cited relative to the number of such papers in the field[60–62]. Here we sought to proactively consider choosing references that reflect the diversity of the field in thought, form of contribution, gender, race, geographic location, and other factors. We used open-source code[63] that uses an automatic classification of gender (gender-api.com) based on the first names of the first and last authors, with possible combinations including man-man, man-woman, woman-man, woman-woman. Excluding self-citations to the first and senior authors of our current paper, the references contain 57% (n=25) man-man, 16% (n=7) man-woman, 16% (n=7) woman-man, and 11% (n=5) woman-woman. Expected proportions in the top 5 neuroscience journals, as reported in Dworkin et al[62] are, respectively: 58.4%, 9.4%, 25.5%, and 6.7%. We look forward to future work that could help us to better understand how to support equitable practices in science.

1. Betzel, R. F. *et al.* The community structure of functional brain networks exhibits scale-specific patterns of inter- and intra-subject variability. *Neuroimage* (2019) doi:10.1016/j.neuroimage.2019.07.003.

2. Bertolero, M. A., Yeo, B. T. T., Bassett, D. S. & D'Esposito, M. A mechanistic model of connector hubs, modularity and cognition. *Nat Hum Behav* **2**, 765–777 (2018).

3. Bertolero, M. A., Yeo, B. T. T. & D'Esposito, M. The diverse club. *Nat Commun* **8**, 1277 (2017).

4. Bertolero, M. A., Yeo, T. B. & D'Esposito, M. The modular and integrative functional architecture of the human brain. *Proceedings of the National Academy of Sciences* **112**, E6798–E6807 (2015).

5. Bassett, D. S., Zurn, P. & Gold, J. I. On the nature and use of models in network neuroscience. *Nature Reviews Neuroscience* **19**, 566–578 (2018).

6. Bassett, D. S. & Sporns, O. Network neuroscience. *Nature Neuroscience* **20**, 353 (2017).

7. Bassett, D. S., Yang, M., Wymbs, N. F. & Grafton, S. T. Learning-induced autonomy of sensorimotor systems. *Nature Neuroscience* **18**, 744–751 (2015).

8. Bassett, D. S. *et al.* Dynamic reconfiguration of human brain networks during learning. *Proceedings of the National Academy of Sciences* **108**, 7641–7646 (2011).

9. Shen, X. *et al.* Using connectome-based predictive modeling to predict individual behavior from brain connectivity. *Nature Protocols* **12**, 506–518 (2017).

10. Finn, E. S. *et al.* Functional connectome fingerprinting: identifying individuals using patterns of brain connectivity. *Nature Neuroscience* **18**, 1664–1671 (2015).

11. Schaefer, A. *et al.* Local-Global Parcellation of the Human Cerebral Cortex from Intrinsic Functional Connectivity MRI. *Cereb Cortex New York N Y 1991* **28**, 3095–3114 (2018).

12. Yeo, B. T. T. *et al.* Functional Specialization and Flexibility in Human Association Cortex. *Cereb Cortex* **26**, 465–465 (2016).

13. Yeo, B. T. T. *et al.* The organization of the human cerebral cortex estimated by intrinsic functional connectivity. *J Neurophysiol* **106**, 1125–65 (2011).

14. Krienen, F. M., Yeo, B. T. T., Ge, T., Buckner, R. L. & Sherwood, C. C. Transcriptional profiles of supragranular-enriched genes associate with corticocortical network architecture in the human brain. *Proc National Acad Sci* **113**, E469–E478 (2016).

15. Buckner, R. L. & Krienen, F. M. The evolution of distributed association networks in the human brain. *Trends in Cognitive Sciences* **17**, 648–665 (2013).

16. Shine, J. M. *et al.* Human cognition involves the dynamic integration of neural activity and neuromodulatory systems. *Nat Neurosci* **22**, 289–296 (2019).

17. Shine, J. M. *et al.* The Dynamics of Functional Brain Networks: Integrated Network States during Cognitive Task Performance. *Neuron* **92**, 544–554 (2016).

18. Bertolero, M. A. & Griffiths, T. L. Is Holism A Problem For Inductive Inference? A Computational Analysis. *Proceedings of the Annual Meeting of the Cognitive Science Society* **36**, (2014).

19. Bertolero, M. A. & Bassett, D. S. On the nature of explanations offered by network science: A perspective from and for practicing neuroscientists. *arXiv preprint arXiv:1911.05031* (2019).

20. Vieira, S., Pinaya, W. H. L. & Mechelli, A. Using deep learning to investigate the neuroimaging correlates of psychiatric and neurological disorders: Methods and applications. *Neurosci Biobehav Rev* **74**, 58–75 (2017).

21. Cornblath, E. J., Lydon-Staley, D. M. & Bassett, D. S. Harnessing networks and machine learning in neuropsychiatric care. *Curr Opin Neurobiol* **55**, 32–39 (2019).

22. Arnemann, K. L. *et al.* Functional brain network modularity predicts response to cognitive training after brain injury. *Neurology* **84**, 1568–74 (2015).

23. Gratton, C., Nomura, E. M., Pérez, F. & D'Esposito, M. Focal Brain Lesions to Critical Locations Cause Widespread Disruption of the Modular Organization of the Brain. *Journal of Cognitive Neuroscience* **24**, 1275–1285 (2012).

24. Gratton, C., Laumann, T. O., Gordon, E. M., Adeyemo, B. & Petersen, S. E. Evidence for Two Independent Factors that Modify Brain Networks to Meet Task Goals. *Cell Reports* **17**, 1276–1288 (2016).

25. Gratton, C. *et al.* Functional Brain Networks Are Dominated by Stable Group and Individual Factors, Not Cognitive or Daily Variation. *Neuron* **98**, 439-452.e5 (2018).

26. Dinga, R. *et al.* Predicting the naturalistic course of depression from a wide range of clinical, psychological, and biological data: a machine learning approach. *Transl Psychiat* **8**, 241 (2018).

27. Tambini, A., Ketz, N. & Davachi, L. Enhanced brain correlations during rest are related to memory for recent experiences. *Neuron* **65**, 280–90 (2010).

28. Tambini, A. & Davachi, L. Persistence of hippocampal multivoxel patterns into postencoding rest is related to memory. *P Natl Acad Sci Usa* **110**, 19591–6 (2013).

29. Hermans, E. J. *et al.* Persistence of Amygdala–Hippocampal Connectivity and Multi-Voxel Correlation Structures During Awake Rest After Fear Learning Predicts Long-Term Expression of Fear. *Cereb Cortex* bhw145 (2016) doi:10.1093/cercor/bhw145.

30. Davidow, J. Y., Foerde, K., Galván, A. & Shohamy, D. An Upside to Reward Sensitivity: The Hippocampus Supports Enhanced Reinforcement Learning in Adolescence. *Neuron* **92**, 93–99 (2016).

31. Gerraty, R. T., Davidow, J. Y., Wimmer, G. E., Kahn, I. & Shohamy, D. Transfer of learning relates to intrinsic connectivity between hippocampus, ventromedial prefrontal cortex, and large-scale networks. *J Neurosci Official J Soc Neurosci* **34**, 11297–303 (2014).

32. Tompary, A., Duncan, K. & Davachi, L. Consolidation of Associative and Item Memory Is Related to Post-Encoding Functional Connectivity between the Ventral Tegmental Area and Different Medial Temporal Lobe Subregions during an Unrelated Task. *J Neurosci Official J Soc Neurosci* **35**, 7326–31 (2015).

33. Silver, D. *et al.* Mastering the game of Go with deep neural networks and tree search. *Nature* **529**, 484–489 (2016).

34. He, T. *et al.* Deep neural networks and kernel regression achieve comparable accuracies for functional connectivity prediction of behavior and demographics. *Neuroimage* 116276 (2019) doi:10.1016/j.neuroimage.2019.116276.

35. Dadi, K. *et al.* Benchmarking functional connectome-based predictive models for resting-state fMRI. *Neuroimage* **192**, 115–134 (2019).

36. Suk, H.-I., Lee, S.-W., Shen, D. & Initiative, A. D. N. Latent feature representation with stacked auto-encoder for AD/MCI diagnosis. *Brain Struct Funct* **220**, 841–859 (2015).

37. Montavon, G., Samek, W. & Müller, K.-R. Methods for interpreting and understanding deep neural networks. *Digit Signal Process* **73**, 1–15 (2018).

38. Samek, W., Wiegand, T. & Müller, K.-R. Explainable Artificial Intelligence: Understanding, Visualizing and Interpreting Deep Learning Models. (2017).

39. Bastian, M., Heymann, S. & Jacomy, M. Gephi: An Open Source Software for Exploring and Manipulating Networks. in *International AAAI Conference on Weblogs and Social Media* (2009).

40. Meo, P. D., Ferrara, E., Fiumara, G. & Provetti, A. Generalized Louvain method for community detection in large networks. *2011 11th Int Conf Intelligent Syst Des Appl* 88–93 (2011) doi:10.1109/isda.2011.6121636.

41. Mucha, P. J., Richardson, T., Macon, K., Porter, M. A. & Onnela, J.-P. Community Structure in Time-Dependent, Multiscale, and Multiplex Networks. *Science* **328**, 876–878 (2010).

42. Ardesch, D. J. *et al.* Evolutionary expansion of connectivity between multimodal association areas in the human brain compared with chimpanzees. *Proc National Acad Sci* **116**, 201818512 (2019).

43. Wei, Y. *et al.* Genetic mapping and evolutionary analysis of human-expanded cognitive networks. *Nat Commun* **10**, 4839 (2019).

44. Hill, J. *et al.* Similar patterns of cortical expansion during human development and evolution. *Proc National Acad Sci* **107**, 13135–13140 (2010).

45. Hwang, K., Bertolero, M. A., Liu, W. B. & D'Esposito, M. The Human Thalamus Is an Integrative Hub for Functional Brain Networks. *J Neurosci* **37**, 5594–5607 (2017).

46. Warren, D. E. *et al.* Network measures predict neuropsychological outcome after brain injury. *Proceedings of the National Academy of Sciences* **111**, 14247–14252 (2014).

47. Betzel, R. F., Bertolero, M. A. & Bassett, D. S. Non-assortative community structure in resting and task-evoked functional brain networks. *Biorxiv* 355016 (2018) doi:10.1101/355016.

48. Kashtan, N. & Alon, U. Spontaneous evolution of modularity and network motifs. *P Natl Acad Sci Usa* **102**, 13773–13778 (2005).

49. Rodriguez, N., Izquierdo, E. & Ahn, Y.-Y. Optimal modularity and memory capacity of neural reservoirs. *Netw Neurosci* **3**, 551–566 (2019).

50. Maglanoc, L. A. *et al.* Predicting cognitive and mental health traits and their polygenic architecture using large-scale brain connectomics. *Biorxiv* 609586 (2019) doi:10.1101/609586.

51. Drysdale, A. T. *et al.* Resting-state connectivity biomarkers define neurophysiological subtypes of depression. *Nature Medicine* **23**, nm.4246 (2016).

52. Baker, J. T. *et al.* Functional connectomics of affective and psychotic pathology. *Biorxiv* 489377 (2018) doi:10.1101/489377.

53. Smith, S. M. *et al.* A positive-negative mode of population covariation links brain connectivity, demographics and behavior. *Nat Neurosci* **18**, 1565–1567 (2015).

54. Essen, D. C. V. *et al.* The WU-Minn Human Connectome Project: An overview. *Neuroimage* **80**, 62–79 (2013).

55. Alfaro-Almagro, F. *et al.* Image processing and Quality Control for the first 10,000 brain imaging datasets from UK Biobank. *Neuroimage* **166**, 400–424 (2018).

56. Miller, K. L. *et al.* Multimodal population brain imaging in the UK Biobank prospective epidemiological study. *Nat Neurosci* **19**, 1523–1536 (2016).

57. Glasser, M. F. *et al.* The minimal preprocessing pipelines for the Human Connectome Project. *Neuroimage* **80**, 105–24 (2013).

58. Griffanti, L. *et al.* ICA-based artefact removal and accelerated fMRI acquisition for improved resting state network imaging. *Neuroimage* **95**, 232–47 (2014).

59. Salimi-Khorshidi, G. *et al.* Automatic denoising of functional MRI data: combining independent component analysis and hierarchical fusion of classifiers. *Neuroimage* **90**, 449–68 (2014).

60. Maliniak, D., Powers, R. & Walter, B. F. The Gender Citation Gap in International Relations. *Int Organ* **67**, 889–922 (2013).

61. Caplar, N., Tacchella, S. & Birrer, S. Quantitative Evaluation of Gender Bias in Astronomical Publications from Citation Counts. *Nat Astronomy* **1**, 0141 (2016).

62. Dworkin, J. D. *et al.* The extent and drivers of gender imbalance in neuroscience reference lists. *Biorxiv* 2020.01.03.894378 (2020) doi:10.1101/2020.01.03.894378.

63. Zhou, D. *et al.* Gender Diversity Statement and Code Notebook v1.0. (2020) doi:10.5281/zenodo.3672110.